\documentclass[10pt,twocolumn,showpacs,longbibliography,amsmath,amssymb,osajnl,floatfix,superscriptaddress]{revtex4-1}

\usepackage{graphicx}
\usepackage{dcolumn}
\usepackage{bm}
\usepackage{color}
\usepackage{txfonts}
\usepackage{microtype}
\usepackage{url}
\usepackage{epstopdf}
\usepackage{indentfirst}
\usepackage{textcomp}
\usepackage{mathrsfs}

\begin{document}

\title{Trapping and acceleration of spin-polarized positrons from $\gamma$ photon splitting in wakefields}

\author{Wei-Yuan Liu}
\thanks{These authors contributed equally to this work.}
\affiliation{Key Laboratory for Laser Plasmas (MOE), School of Physics and Astronomy, Shanghai Jiao Tong University, Shanghai, 200240, China}
\affiliation{Collaborative Innovation Center of IFSA (CICIFSA), Shanghai Jiao Tong University, Shanghai 200240, China}

\author{Kun Xue}
\thanks{These authors contributed equally to this work.}
\affiliation{Key Laboratory for Nonequilibrium Synthesis and Modulation of Condensed Matter (MOE), School of Physics, Xi'an Jiaotong University, Xi'an 710049, China}

\author{Feng Wan}	
\affiliation{Key Laboratory for Nonequilibrium Synthesis and Modulation of Condensed Matter (MOE), School of Physics, Xi'an Jiaotong University, Xi'an 710049, China}	

\author{Min Chen}
\email{minchen@sjtu.edu.cn}
\affiliation{Key Laboratory for Laser Plasmas (MOE), School of Physics and Astronomy, Shanghai Jiao Tong University, Shanghai, 200240, China}
\affiliation{Collaborative Innovation Center of IFSA (CICIFSA), Shanghai Jiao Tong University, Shanghai 200240, China}

\author{Jian-Xing Li}
\email{jianxing@xjtu.edu.cn}
\affiliation{Key Laboratory for Nonequilibrium Synthesis and Modulation of Condensed Matter (MOE), School of Physics, Xi'an Jiaotong University, Xi'an 710049, China}

\author{Feng Liu}
\affiliation{Key Laboratory for Laser Plasmas (MOE), School of Physics and Astronomy, Shanghai Jiao Tong University, Shanghai, 200240, China}
\affiliation{Collaborative Innovation Center of IFSA (CICIFSA), Shanghai Jiao Tong University, Shanghai 200240, China}

\author{Su-Ming Weng}
\affiliation{Key Laboratory for Laser Plasmas (MOE), School of Physics and Astronomy, Shanghai Jiao Tong University, Shanghai, 200240, China}
\affiliation{Collaborative Innovation Center of IFSA (CICIFSA), Shanghai Jiao Tong University, Shanghai 200240, China}

\author{Zheng-Ming Sheng}
\affiliation{Key Laboratory for Laser Plasmas (MOE), School of Physics and Astronomy, Shanghai Jiao Tong University, Shanghai, 200240, China}
\affiliation{Collaborative Innovation Center of IFSA (CICIFSA), Shanghai Jiao Tong University, Shanghai 200240, China}
\affiliation{SUPA, Department of Physics, University of Strathclyde, Glasgow G40NG, UK}

\author{Jie Zhang}
\affiliation{Key Laboratory for Laser Plasmas (MOE), School of Physics and Astronomy, Shanghai Jiao Tong University, Shanghai, 200240, China}
\affiliation{Collaborative Innovation Center of IFSA (CICIFSA), Shanghai Jiao Tong University, Shanghai 200240, China}

\date{\today}

\begin{abstract}
Energetic spin-polarized positrons are extremely demanded for forefront researches, such as $e^- e^+$ collider physics, but making compact positron sources is still very challenging. Here we put forward an efficient scheme of trapping and acceleration of polarized positrons in plasma wakefields. Seed electrons colliding with a bichromatic laser create polarized $\gamma$ photons which then split into $e^- e^+$ pairs via nonlinear Breit-Wheeler process with an average (partial) positron polarization above 30\% (70\%). Over 70\% positrons are trapped and accelerated in recovered wakefields driven by a hollow electron beam, obtaining an energy gain of 3.5 GeV/cm with slight depolarization. This scheme provides a potential for constructing compact and economical positron sources for future applications.
\end{abstract}

\maketitle

Plasma-based wakefield accelerators have attracted worldwide attentions in recent years due to their capability of providing acceleration gradients three orders of magnitude higher than conventional radio-frequency accelerators \cite{tajima1979,chen1985,nakajima1995}. Over the past decades, the wakefield acceleration of electrons has been developed rapidly \cite{esarey2009,downer2018}. This promises a new possibility for future electron-positron ($e^- e^+$) colliders with relatively compact size and low cost \cite{leemans2009,schroeder2010,nakajima2019}. To this end, trapping and acceleration of polarized positron beams are highly demanded on top of the advantage of high acceleration gradient for electrons in wakefields \cite{artru2008,clarke2008}. However, generation, polarization, trapping and acceleration of such positron beams in plasma wakefields are still quite challenging.

Although plenty of schemes \cite{esarey2009} have been proposed and studied for the effective trapping and acceleration of electrons in plasma wakefields \cite{pukhov2002,lu2006}, those schemes are not applicable for positrons since the transverse fields in nonlinear wakes usually defocus positrons, which makes continuous positron acceleration impossible. To overcome this issue, an amount of theoretical schemes have been proposed to simultaneously accelerate and focus positrons by using special driver or plasma structures, such as Laguerre-Gaussian laser pulses \cite{vieira2014}, hollow electron beam drivers \cite{jain2015}, and  finite-radius plasma columns \cite{diederichs2019}. But, unfortunately, in  those studies the generation and injection of positrons have to be pre-provided. In recent FACET experiments, the positron accelerations have been demonstrated  to run in a self-loaded plasma wakefield \cite{corde2015}  or  a hollow plasma channel \cite{gessner2016,yakimenko2016}. However, a pre-accelerated relativistic positron beam is also required and the beam polarization has not been studied yet.

Positrons are commonly polarized either via radiative process (Sokolov-Ternov effect) in a storage ring \cite{sokolov1964, baier1967, baier1972} or via high-energy polarized $\gamma$ photons interacting with a high $Z$-target (Bethe-Heitler pair production) \cite{variola2014}. For the former the polarization time is rather long since the magnetic fields of a synchrotron are quite weak; for the latter the positron density is limited by the low photon luminosity \cite{omori_2006,scott2011,abbott2016}. Recently, the state-of-the-art laser pulses with peak intensities up to $10^{22}$ W/cm$^2$ \cite{gales2018,shen2018,Yoon2019,danson2019} enable to excite nonlinear quantum electrodynamics (QED) processes \cite{ritus1985,xie2017} in laser-matter interaction \cite{piazza2012,Sarri2015,
Cole2018,Poder2018}. And, polarized GeV-level positron beams can be created via employing asymmetric spin-resolved probabilities of nonlinear Breit-Wheeler (BW) pair production in a bichromatic \cite{chen2019} or elliptically-polarized laser pulse \cite{wan2020} (transverse polarization; the polarization of intermediate photons was not considered therein), or via the helicity transfer from polarized electrons (longitudinal polarization) \cite{li2020_2}. However, in those methods the positron energies are limited by those of the scattering electrons via intermediate photons and impossible to achieve the level of hundreds of GeVs, and the beam qualities, such as the energy spread and emittance, are far worse than those of the beams from conventional accelerators, which severely restrict the applications in high-energy and particle physics (e.g. the polarized $e^- e^+$ collider \cite{clarke2008}).

\begin{figure}[t]		
\includegraphics[width=0.95\linewidth]{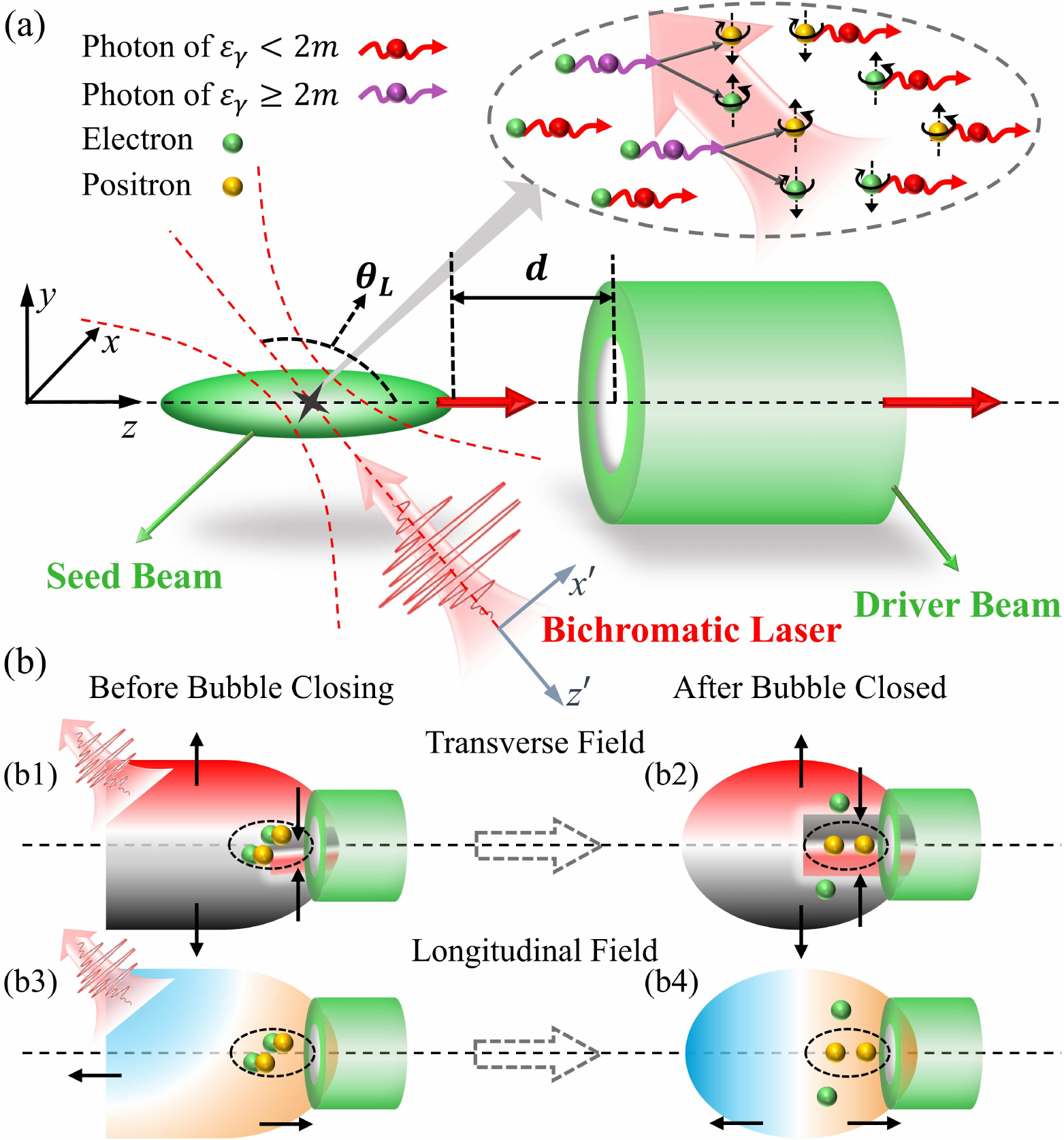}\\
	\caption{Interaction scenario of polarization, trapping and acceleration of positrons. (a) A hollow electron beam copropagates with a seed electron beam along $+z$ direction with a separation distance $d$. A LP bichromatic laser pulse, polarizing in $x'-z'$ plane, collides with the seed beam with a collision angle $\theta_L$. (b) When the laser leaves, the bubble gradually recovers and traps the positrons. During the bubble closing, [(b1) and (b2)] the transverse fields (red-black gradients) near the bubble axis can focus the positrons and repel the electrons, and the outer fields act reversely; [(b3) and (b4)] the front part of the longitudinal fields (orange-blue gradients) accelerate the positrons and decelerate the electrons. The black arrows indicate the force felt by the positrons due to the wakefields.}
	\label{fig1}
\end{figure}

\begin{figure}[t]
	\includegraphics[width=1.0\linewidth]{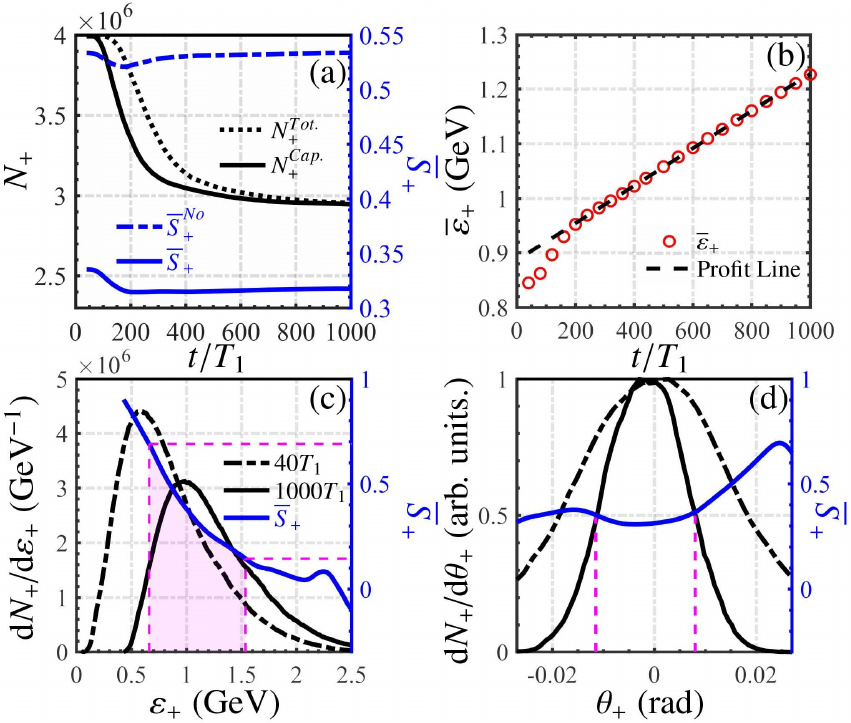}

		\caption{Evolutions of the total positron number $N_+^{Tot.}$ inside the first bubble (black-dotted), captured positron number inside the focusing region $N_+^{Cap.}$ (black-solid) and average polarization of captured positrons $\overline{S}_+$ (blue-solid) with considering the polarization of intermediate $\gamma$ photons, respectively. The blue-dash-dotted curve (``$\overline{S}_+^{No}$'') indicates the average polarization of captured positrons for the case of artificially neglecting the polarization of intermediate $\gamma$ photons. (b) Average energy of captured positrons $\overline{\varepsilon}_+$ (red-circles) and its linear profit (black-dashed) vs the interaction time $t$, respectively, with an acceleration gradient $G\approx3.58$ GV/cm. (c) Energy spectra of captured positrons d$N_+$/d$\varepsilon_+$ [at the instant of finishing of pair creation $t_i=40T_1$ (black-dash-dotted) and at the end of the simulation $t_f=1000T_1$ (black-solid)] and $\overline{S}_+$ at $t_f$ (blue-solid) vs the positron energy $\varepsilon_+$, respectively. (d) Normalized angular distributions of positrons  at $t_i$ (black-dash-dotted) and $t_f$ (black-solid), and $\overline{S}_+$ at $t_f$ (blue-solid) vs the transverse angular divergence of the positrons  $\theta_+=$arctan$(p_{+,x}/p_{+,z})$, respectively. Initial parameters of the laser pulse, electron beams and plasma are given in the text.}
		\label{fig2}
\end{figure}

In this Letter, we propose  a compact scheme to generate polarized positrons and inject them into plasma wakefields with further acceleration to high energies. The positron generation and polarization are studied quantum mechanically,  while the bubble-recovery-based positron trapping and following acceleration and depolarization in wakefields semi-classically. The interaction schematic is shown in Fig.~\ref{fig1}. A hollow electron beam working as a wake driver propagates into a low-density plasma and excites nonlinear wakefields (bubbles). Behind it, another copropagating seed electron beam collides with an ultra-intense linearly-polarized (LP) bichromatic laser pulse to emit abundant LP $\gamma$ photons via nonlinear Compton scattering, which could further decay into transversely polarized pairs through nonlinear BW process [see Fig.~\ref{fig1}(a)] due to the asymmetric pair production and polarization probabilities in the laser positive and negative half cycles. We underline that in this study the polarization of intermediate $\gamma$ photons has been taken into account, otherwise the yield and polarization of the positrons will be remarkably overestimated. During the collision of the laser and seed beam, the wake structure driven by the driver beam is first destroyed and then gradually self-recovers at the downstream of the laser-seed-beam collision point.  Some of the  created high-energy polarized positrons can be trapped in the recovered wakefields [see Figs.~\ref{fig1}(b1) and (b2)] and  then accelerated by the wakefields [see Figs.~\ref{fig1}(b3) and (b4)]. In our simulations over 70\% positrons are finally injected into the wake and get further acceleration to an average energy beyond 1.2 GeV in 1 millimeter, with an average polarization exceeding 30\%. The partial polarization of the positrons within the full width at half maximum (FWHM) of the energy spectrum can exceed 70\% [see Fig.~\ref{fig2}(c)]. The detailed injection and acceleration processes are discussed in the following.

 We develope  a Monte Carlo algorithm and implement it into two-dimensional QED particle-in-cell (PIC) code (benchmarked by EPOCH code \cite{Arber2015}) to describe the creation and polarization of the pairs quantum mechanically by using spin-resolved probabilities of nonlinear BW pair production \cite{supplemental}, which are derived from the QED operator method \cite{baier1973} in the local constant field approximation (valid at the invariant laser field parameter $a_0=|e| E_0/m\omega\gg1$) \cite{ritus1985, baier1998, ilderton2019,piazza2019}. To efficiently generate $\gamma$ photons and pairs requires the nonlinear QED parameters $\chi_e\equiv|e|\sqrt{-(F_{\mu v}p_e^v)^2}/m^3\gtrsim1$ (for electrons) and $\chi_{\gamma}\equiv|e|\sqrt{-(F_{\mu v}k_{\gamma}^v)^2}/m^3\gtrsim1$ (for $\gamma$ photons) \cite{ritus1985,baier1998}. Here, $F_{\mu v}$ is the field tensor, $p_e^v$ and $k_{\gamma}^v$ the 4-momenta of electron and $\gamma$ photon, respectively, $e$ and $m$ the electron charge and mass, respectively, and $E_0$ and $\omega$ the laser amplitude and frequency, respectively. Relativistic units with $c=\hbar=1$ are used throughout. The simulations of spin-resolved electron (positron) dynamics and photon emission and polarization follow the semi-classical algorithms in Refs.~\cite{li2019,li2020,wan2020}. See more details of our simulation method in \cite{supplemental}.

The simulation parameters of the laser pulse, electron beams and plasma are summarized as follows. A tightly-focused LP Gaussian bichromatic laser pulse propagates along $-z'$ direction with $\theta_L=105^{\circ}$ and polarizes in $x'-z'$ plane, with wavelengths $\lambda_1=1 \mu$m (period $T_1$) and $\lambda_2=0.5 \mu$m, pulse durations $\tau_1=\tau_2=6T_1$, focal radii $w_1=w_2=2 \mu$m, and peak amplitudes $a_1=4a_2\approx67$ (corresponding to the peak intensities $I_1=4I_2\approx6.15\times10^{21}$ W/cm$^2$). An unpolarized elliptical seed beam propagates along $+z$ direction, with an average energy $\varepsilon_{s,0}=4$ GeV, major axis $L_{maj}=7\mu$m, and minor axis $L_{min}=2\mu$m. A hollow driver beam is initially placed at the entrance of the plasma, with an average energy $\varepsilon_{d,0}=1$ GeV, outer radius $w_{out}=3\mu$m, inner radius $w_{in}=1.5\mu$m, and length $L_h=9\mu$m. The density, energy spread and angular divergence of the two electron beams are  $n_{s,0}=n_{d,0}=0.1n_c$ with a Gaussian distribution (the critical density $n_c=1.1\times10^{21}$ cm$^{-3}$ with respect to the laser pulse with wavelength of $\lambda_1$), $\Delta\varepsilon_{s,0}/\varepsilon_{s,0}=\Delta\varepsilon_{d,0}/\varepsilon_{d,0}=0.1$, and $\Delta \theta_{s}=\Delta \theta_{d}=0.1$ mrad, respectively. Here the delay distance of the two electron beams is $d=0\mu$m. More parameter scans for the driver beam size and other effects are shown in Fig.~\ref{fig4}. The density of the background plasma (composed of $H^+$ and electrons) is $n_{p,0}=0.01n_c$. Note that the efficient excitation of a wakefield with central focusing fields for the positrons requires the driver beam satisfying $w_{in}/\sigma_x\geq 3$ and $k_p\sigma_z\leq2$ \cite{jain2015,rosenweig2004}, where $\sigma_x$ and $\sigma_z$ are the transverse and longitudinal sizes of the driver beam and $k_p=2\pi/\lambda_p$ with $\lambda_p=\sqrt{\pi m/n_{p,0}e^2}$. Here we use $w_{in}/\sigma_x=3$ and $k_p\sigma_z\approx 1.8$, and the simulation domain is $60\lambda_1 (x)\times80\lambda_1 (z)$ with grid resolutions d$x$ = d$z=\lambda_1/50$.

\begin{figure}[t]
	\includegraphics[width=1.0\linewidth]{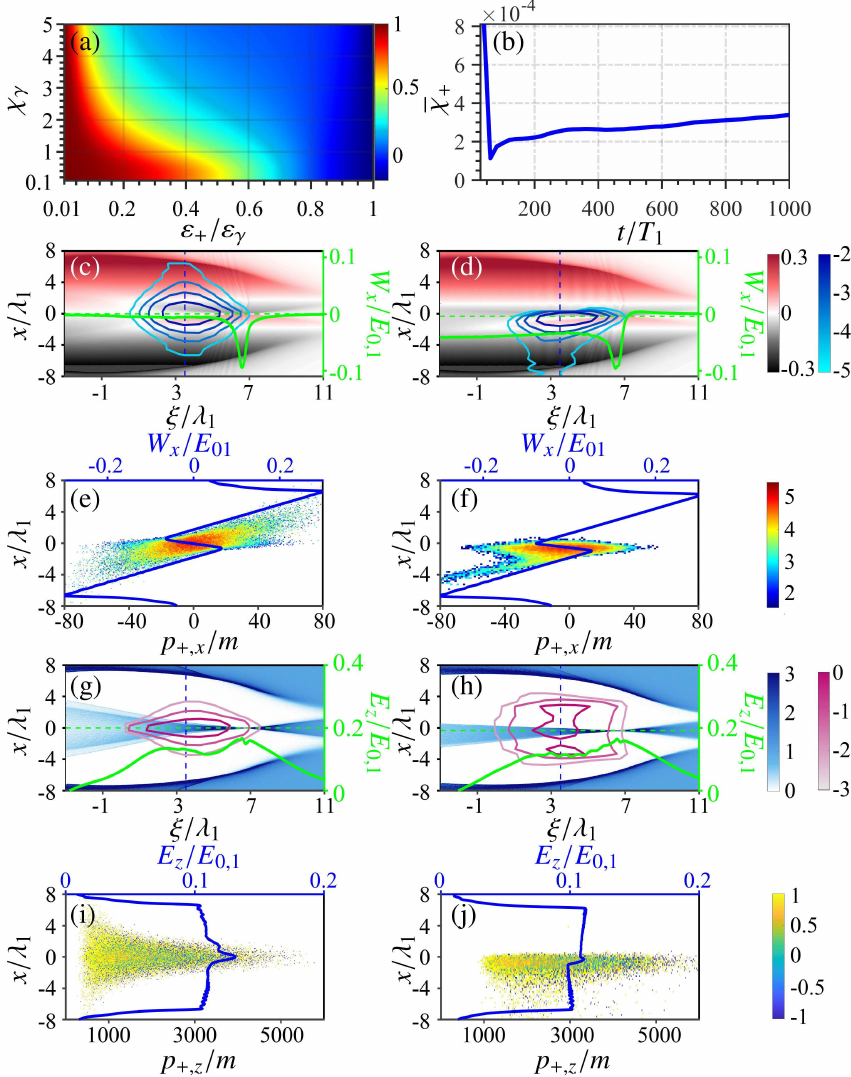}
\caption{(a) Analytical value of $\overline{S}_+$ with respect to $\chi_{\gamma}$ and $\varepsilon_+/\varepsilon_{\gamma}$; see the analytical expression in \cite{supplemental}. (b) $\overline{\chi}_+$ vs $t$. (c) [(d)] Focusing fields $W_x=E_x-B_y$ felt by positrons normalized by $E_{0,1}=2\pi m /\lambda_1 e$ (red-black gradients) and positron density contours log$_{10}(n_+/n_{p,0})$ (cyan-blue gradients) with respect to the co-moving frame variable $\xi=z-t$ and $x$. The green curve represents the longitudinal distribution of $W_x$ (with $x=x_{wake}$+d$x$ and $t=100T_1$ [$t=1000T_1$], where the bubble axis is at $x_{wake}\approx-0.03\lambda_1$ [$x_{wake}\approx-0.34\lambda_1$]). $E_x$ and $B_y$ are transverse components of the wakefields, respectively. (e) [(f)] Distribution of the positron number (log$_{10}N_+$) with respect to $p_{+,x}$ and $x$. The blue line represents $W_x$ at $\xi=3.5\lambda_1$ vs $x$ at $t=100T_1$ [$t=1000T_1$]. (g) [(h)] Distributions of background electron density $n_p/n_{p,0}$ (white-blue gradients) and contours of seed electron density log$_{10}(n_s/n_{p,0})$ (white-magnet gradients) with respect to $\xi$ and $x$. The green curve represents the longitudinal electric field $E_z$ at $x=x_{wake}$+d$x$ vs $\xi$ at $t=100T_1$ [$t=1000T_1$]. (i) [(j)] Distributions of $\overline{S}_+$ with respect to $p_{+,z}$ and $x$. The blue curve represents $E_z$ at $\xi=3.5\lambda_1$ vs $x$, at $t=100T_1$ [$t=1000T_1$]. Other parameters are the same with those in Fig.~\ref{fig2}.}

\label{fig3}
\end{figure}

The main results of the positron trapping, acceleration and polarization are shown in Figs.~\ref{fig2} and \ref{fig3}. The pair production process is completed at the distance of $t_i\approx40T_1$, where the bubble has not fully recovered yet, and nearly $4\times10^6$ positrons are created with a yield ratio $N_+/N_s\approx0.4\%$ (corresponding to a density of $n_+\sim10^{-4}n_c$) and an average polarization (mainly along the magnetic field direction $y$) $\overline{S}_+\approx33.52\%$ [see Fig.~\ref{fig2}(a)]. As we mentioned before, if the polarization of intermediate $\gamma$ photons is artificially neglected as usual, $\overline{S}_+$ will be considerably overestimated by exceeding 68\% [see the blue-dash-dotted line in Fig.~\ref{fig2}(a), $\overline{S}_+^{No}\approx$ 53.5\% at $t_f$]. Therefore we include this effects in our simulations and the analytical calculation of positron polarization is shown in Fig.~\ref{fig3}(a). The polarization degree is inversely proportional to the positron energy, which affects the final polarization distribution of the accelerated positrons [see Fig.~\ref{fig2}(c)]. Since in our case the QED parameter of the positron $\chi_+\propto a_{wake}\gamma_+[1-{\rm cos}(\theta_L)]\ll1$ [see Fig.~\ref{fig3}(b)], the radiative depolarization effect is very weak, where $a_{wake}$ represents the invariant field parameter and $\gamma_+$ the Lorentz factor of positron.
The depolarization effect derived from the spin procession in the wakefield govern by the Thomas-Bargmann-Michel-Telegdi equation \cite{thomas1926,thomas1927,bargmann1959} is also quite weak \cite{vieira2011,wen2019}. Consequently, the final positron polarization distribution mainly depends on the initial pair creation process and the conditions for the positron selection during the trapping and continuous acceleration processes. The last two processes rely on the wakefield structure. The positrons inherit transverse momenta $p_{+,x}$ from the seed electrons via intermediate $\gamma$ photons. During acceleration, the positrons with large-$p_{+,x}$ may escape from the central focusing region and are then expelled out of the bubble by the outer defocusing transverse field. The positrons with low-$p_{+,x}$ can be continuously trapped in the acceleration phase [see Figs.~\ref{fig3}(c)-(f)]. Finally at $t_f=1000T_1$ about 74.12\% positrons are accelerated with an average energy increase of about 350 MeV in a distance $\lesssim 1$ mm, reaching $\overline{\varepsilon}_+\approx1.24$ GeV [see Fig.~\ref{fig2}(b)], and the acceleration gradient is $G\approx3.53$ GV/cm [see Figs.~\ref{fig3}(g) and (h)]. The final average positron polarization is $\overline{S}_+\approx31.77\%$, which is only slightly depolarized after acceleration [see Fig.~\ref{fig2}(a)]. In the period of $200T_1\lesssim t\lesssim1000T_1$ some high-energy positrons with low polarization gradually escape from the focusing region [see Fig.~\ref{fig3}(j)], therefore, the polarization increases a little. At $t_f$ the positron polarization distribution around the peak area of the energy spectrum within the FWHM declines approximately from 70\% to 15\%. Such distribution provides a possible way to further increase the polarization by the energy-selection technique \cite{alexander2009}.

Besides the trapping ratio and polarization degree, the energy spread and divergence are also important factors for future applications. In Fig.~\ref{fig2}(c), we find that the relative energy spread of the positrons decreases by about 26\% after the wake acceleration compared to the instant of the pair creation $t_i$. While, the absolute energy spread does not increase during the acceleration, because the seed beam not only `provides' the pairs but also flattens the local acceleration field [see Figs.~\ref{fig3}(g)-(j)] assuring uniform acceleration and avoiding energy dispersion. The angular divergence of the positron beam is also improved by the focusing field to $\Delta\theta_+\approx20$ mrad, which is about 50\% lower than that at $t_i$ [see Fig.~\ref{fig2}(d), the polarization is nearly uniform ($\overline{S}_+\approx32.67\%$) within the FWHM labelled by the two dashed purple lines]. One can see that there is an asymmetric angular distribution at $t_f$ in Fig.~\ref{fig2}(d). This is induced by the unbalanced plasma perturbations [indicated in Figs.~\ref{fig3}(d) and (f)], originating from the laser incidence from one side.

\begin{figure}[t]
	\includegraphics[width=1.0\linewidth]{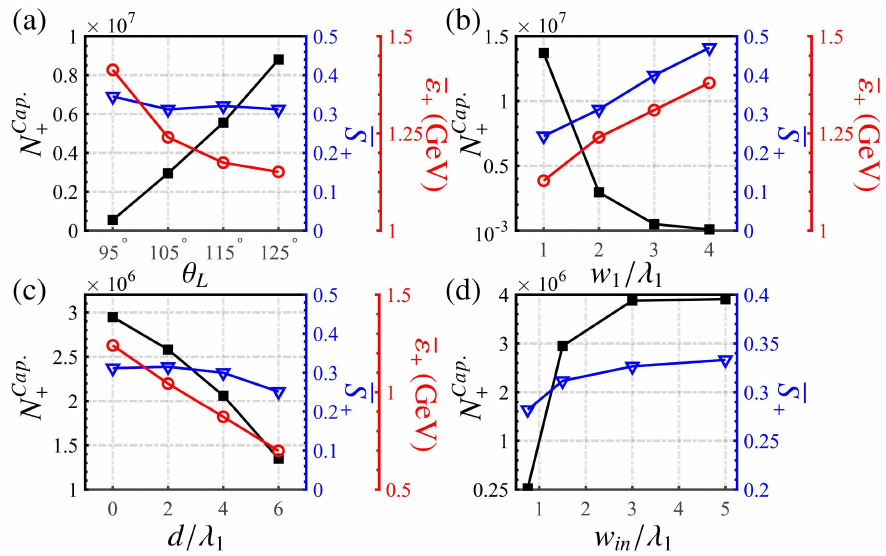}
		\caption{(a)-(d): Variations of $N_+^{Cap.}$ (black, square-mark), $\overline{S}_+$ (blue, triangle-mark) and $\overline{\varepsilon}_+$ (red, circle-mark) of captured positrons at $t_f$ with respect to $\theta_{L}$, $w_1$ ($w_1=w_2$ with a fixed laser energy), $d$ and $w_{in}$ respectively. Other parameters are the same with those in Fig.~\ref{fig2}.}
		\label{fig4}
\end{figure}

Finally, we study the impact of the initial parameters on the trapping, acceleration and polarization of the positrons in Fig.~\ref{fig4}.
As the collision angle $\theta_L$ increases from $95^{\circ}$ to $125^{\circ}$ [see the interaction scenario in Fig.~\ref{fig1}], the probabilities of photon emission and pair production (determined by $\chi_e \propto a_0 \gamma_e$[1-cos($\theta_L$)] and $\chi_\gamma \propto a_0 k_\gamma$[1-cos($\theta_L$)], respectively) are both enhanced, thus, $N_+^{Cap.}$ increases. At the same time the average energy decreases due to $\overline{\varepsilon}_+\propto  \sum \varepsilon_\gamma/N_+^{Cap.}\propto \sum\varepsilon_{s,0}/N_+^{Cap.}$. $\overline{S}_+$ decreases as well since the asymmetry of the spin-resolved pair production probabilities in the laser positive and negative half cycles is weakened and the radiative depolarization effect is enhanced [see Fig.~\ref{fig4}(a)].
As the laser focal radius $w_1 (w_2)$ increases with a fixed laser energy $J$, the laser peak amplitude $a_0\propto\sqrt{J}/w_1$ and $N_+^{Cap.}\propto a_0^2\propto J/w_1^2$ decrease, and accordingly, $\overline{S}_+$ and $\overline{\varepsilon}_+$ are both enhanced [see Fig.~\ref{fig4}(b)].
As the distance between the seed and driver beams $d$ rises up, more low-energy positrons with high polarization cannot enter the focusing region to be steadily accelerated (i.e. $N_+^{Cap.}$ and $\overline{S}_+$ both decrease), and the enhancement of the acceleration field by the seed beam [indicated in Figs.~\ref{fig3}(g) and (h)] is weakened (i.e. $\overline{\varepsilon}_+$ decreases) [see Fig.~\ref{fig4}(c)]. Thus, the condition of $d\lesssim \lambda_p/2$ should be satisfied.
As the inner radius of the driver beam $w_{in}$  increases (more feasible in experiments), more plasma electrons converge into the hollow region to create a larger transverse size of the focusing region \cite{jain2015,supplemental}, therefore, more positrons can be trapped and the depolarization effect induced by the escape of high-polarization positrons is weakened (i.e. $\overline{S}_+$ increases) [see Fig.~\ref{fig4}(d) and more details in Ref.~\cite{supplemental}].
We underline that in our scheme the positrons are transversely polarized and can be used to investigate specific Triple Gauge Couplings and $W$-physics, and to test the validity of the standard model and discovering new physics \cite{fleischer1994}. In further, the arbitrary spin orientation can be realized through a proper spin rotator \cite{moortgatpick2008}.

In conclusion, utilizing both advantages of laser-driven QED process and plasma wakefield acceleration we have proposed a compact scheme for positron polarization, trapping and acceleration. Dense GeV positron beams with a spin polarization up to 70\% and improved beam quality compared with the scheme of single laser-electron collision can be achieved. By using multi-staged wakefield acceleration with currently achievable laser facilities, this scheme also provides a possible way to generate highly-polarized positron beams with hundreds of GeVs energy for future compact research and application platforms of high-energy and particle physics.\\

\textbf{Acknowledgement:} This work was supported by National Natural Science Foundation of China (Grants No. 11991074, 11721091, 11874295, 11804269 and 11905169), the National Key R\&D Program of China (Grant No. 2018YFA0404801), and the Science Challenge Project of China (Grants No. TZ2016099 and TZ2018005).

\bibliography{mybib}

\end{document}